%% file: main.tex
\documentclass[11pt,a4paper]{article}
\usepackage[hyperref]{acl2020}
\usepackage{times}
\usepackage{latexsym}
\usepackage{graphicx}

\usepackage{mathtools} 
\usepackage{amsmath} 
\usepackage{multirow}
\usepackage{balance}
\usepackage{subcaption}
\usepackage{url}
\usepackage{xcolor}

\newcommand{\Part}[1]{\noindent\textbf{#1}}
\usepackage{enumitem}
\usepackage{float}
\usepackage{comment}

\usepackage{microtype}

\aclfinalcopy 

\title{COVID-19: Social Media Sentiment Analysis on Reopening}

\author{Mohammed Emtiaz Ahmed, Md Rafiqul Islam Rabin, Farah Naz Chowdhury\\
    Department of Computer Science, University of Houston  \\
\texttt{mahmed24@uh.edu}, \texttt{mrabin@central.uh.edu}, \texttt{fchowdh2@central.uh.edu} }

\date{}

\begin{document}
\maketitle

\input{abstract}

\input{intro}

\input{related}

\input{dataset}

\input{figs2/wordcloud}
\input{methodology}

\input{results}

\input{future}

\input{conclusion}

\balance 

\bibliography{main}
\bibliographystyle{acl_natbib}

\end{document}

%% file: abstract.tex
\begin{abstract}
The novel coronavirus (COVID-19) pandemic is the most talked topic in social media platforms in 2020. People are using social media such as Twitter to express their opinion and share information on a number of issues related to the COVID-19 in this stay at home order. 
In this paper, we investigate the sentiment and emotion of peoples in the U.S.A on the subject of \textit{reopening}. We choose the social media platform Twitter for our analysis and study the Tweets to discover the sentimental perspective, emotional perspective, and triggering words towards the reopening.
During this COVID-19 pandemic, researchers have made some analysis on various social media dataset regarding \textit{lockdown} and \textit{stay at home}. However, in our analysis, we are particularly interested to analyse public sentiment on \textit{reopening}. Our major finding is that when all states resorted to lockdown in March, people showed dominant emotion of fear, but as reopening starts people have less fear. While this may be true, due to this reopening phase daily positive cases are rising compared to the lockdown situation. Overall, people have a less negative sentiment towards the situation of reopening.
\end{abstract}

%% file: intro.tex
\section{Introduction}
COVID-19, a new strain of coronavirus, originated in Wuhan, China in December 2019~\cite{huang2020clinical}. On $11^{th}$ March 2020, the World Health Organization announced the COVID-19 outbreak as a pandemic. This highly infectious COVID-19 has crossed boundaries in a speed anybody could have imagined turning the world upside down. COVID-19 has already claimed more than 0.3 million lives, with around 6 million people infected in more than 200 countries or territories \footnote{\url{https://www.who.int/emergencies/diseases/novel-coronavirus-2019/}} up to May 31, 2020.

It has shaken the global economic and social structure so badly, that resuming normal life is a thing to envision.  As a measure to control the situation, a number of countries have resorted to complete lockdown \cite{mitja2020experts}. When the whole world is COVID-19 stricken, the lockdown has been the only solution to get a hold \cite{ghosal2020impact}. Since then much of the debate has been going on how the countries should ease the lockdown program. As lockdown is not the ultimate solution, easing the restrictions is the only way to minimize economic loss and keep the society functional. As the whole world is still not free from the deadly clutch of corona and continuing to claim human lives, reopen decisions from the government is a huge step to take. Public sentiment towards this reopen decision can help policymakers and general people to better understand the situation through classified information.

This COVID-19 outbreak has not only caused economic breakdown \cite{fernandes2020economic} across the world but also left a great impact on human minds. To capture human emotions, social media websites work as one of the best possible sources \cite{de2012nature}. It is the fastest way for people to express themselves and thus the news feed is flooded with data reflecting thoughts dominating the people’s minds at this time. Apart from sharing their thoughts, people use these media to disseminate the information. During this lockdown, people have taken social networks to express their feelings and thus find a way to calm themselves down.

Emotions and sentiments are the driving force that people are sharing in social media. If we analyze the social media posts then we can find the insights of those posts along with emotions and sentiments. From this kind of analysis, we may find what people like, what they want, and what are the main concerns of them. Twitter has a large number of daily active users all around the world where people share their thoughts and information regarding any topic of recent concern through this medium. It is a searchable archive of human thought. The affluence of publicly available data shared through Twitter has encouraged researchers to determine the sentiments on almost everything, including sentiments towards any product, service. Researchers and practitioners are already giving useful information on issues related to this COVID-19 outbreak. We have mainly contributed to the issue of the reopening phase giving an insight into people's reactions. Coordination of responses from online and from the real-world is sure to reveal promising results to address the current crisis. Our major finding is that during this reopen phase fear among people is less dominant compared to the previous times when the states made lockdown decisions. Another finding is that as the reopen phase starts, the number of new cases is also increasing. People are sharing their thoughts and opinions to combat this situation.


\smallskip
\Part{Contributions.} This paper makes the following contributions.
\begin{itemize} [nosep]
    \item First, we generate the word cloud and N-gram representation of tweets to see the insights of twitter users during the reopening phase (more details in section ~\ref{sec:result_most_freq}).
    \item Second, we conduct a shallow analysis to study the sentiment and the emotion in reopening related tweets (more details in section ~\ref{sec:result_emotion} and section ~\ref{sec:result_sentiment}).
    \item Third, we analyze the actual effects of lockdown and reopening using real data (more details in section ~\ref{sec:result_effect}).
    \item Finally, we have made our code publicly available for an extensible work in similar areas.
\end{itemize}

\smallskip
The structure of this paper is organized as follows.
Section~\ref{sec:related} presents a literature review and some considerations of the previous work.
Section~\ref{sec:dataset} contains the information of dataset.
Section~\ref{sec:methodology} describes the experimental setup to analysis the reopening tweet.
Section~\ref{sec:results} provides an analysis of the results and findings.
Section~\ref{sec:threats} discusses some of the challenges and threats to validity.
Finally, Section~\ref{sec:conclusion} concludes the paper with future works.

%% file: related.tex
\section{Related Work}
\label{sec:related}

Researchers all around the world are analyzing Twitter data to discover people's reactions to the corona virus related issues such as lockdown, safety measures. \cite{848091d31dc84e198935a6111a48cd25} analyzed Twitter data and found that people of India were positive towards the lockdown decision of their government to flatten the curve. \cite{dubey} made a sentiment analysis on the tweets related to COVID19 in some or the other way from  March 11  to March 31 on twelve different countries, USA, Italy, Spain, Germany, China, France, UK, Switzerland, Belgium, Netherland, and Australia. They found that while most of the countries are taking a positive attitude towards this situation, there is also negative emotion such as fear, sadness present among people. Countries specially France, Switzerland, Netherlands, and the United States of America have shown distrust and anger more frequently compared to other countries. \cite{rajput2020word} Made a statistical analysis of Twitter data posted during this Coronavirus outbreak. The number of tweet ids tweeting about coronavirus rose rapidly making several peaks during February and March. An empirical analysis of words in the messages showed that the most frequent words were Coronavirus, Covid19, and Wuhan. The immense number of tweet messages within a period of 2-3 months and the frequency of these words clearly show the threats the global population is exposed to. \cite{info:doi/10.2196/19016}  Aimed to identify the main topics posted by Twitter users related to the COVID-19 pandemic between February 2, 2020, and March 15, 2020. Users on Twitter discussed 12 main topics related to COVID-19 that were grouped into four main themes, 1) the origin of the virus; 2) its sources; 3) its impact on people, countries, and the economy that were represented by six topics: deaths, fear, and stress, travel bans and warnings, economic losses, panic, increased racism, and 4) the last theme was ways of mitigating the risk of infection. \cite{cinelli2020covid19} ran a comparative analysis of five different social media platforms Twitter, Instagram, YouTube, Reddit, and Gab about information diffusion about COVID-19. They wanted to investigate how information is spread during any critical moment. Their analysis suggests that information spread depends on the interaction pattern of users engaged with the topic rather than how reliable the information is.
\cite{newazCovid}  Identified public sentiment using coronavirus specific tweets and provided insights on how fear sentiment evolved over time specially when COVID-19 hit the peak levels in the United States.  As they classified tweets using sentiment analysis, the fear sentiment, which was the most dominant emotion across the entire tweets and by the end of march its seriousness became clearly evident as the fear curve showed a steep rise.

%% file: dataset.tex
\section{Dataset}
\label{sec:dataset}

We used data from multiple sources. The primary source of this analysis is the social platform Twitter. We used the Python library \texttt{Tweepy} \footnote{\url{https://github.com/tweepy/tweepy}} \cite{roesslein2009tweepy} and Twitter developers API \footnote{\url{https://developer.twitter.com/en/docs}} for collecting the tweet data from Twitter. We used a unique Twitter dataset, which is specifically collected for this study using a date range from May 3, 2020, to May 15, 2020. Tweets were extracted using the hashtags: \#covid19, \#covid, \#corona, \#coronaviras, \#corona-virus, \#covid19-virus, and \#sarscov2. We collected a total of 5,703,590 tweets from Twitter. In addition, we collected the COVID-19 time-series dataset from the GitHub repository \href{https://github.com/CSSEGISandData/COVID-19}{CSSEGISandData/COVID-19} ~\cite{dong2020interactive}, maintained by the amazing team at Johns Hopkins University Center for Systems Science and Engineering (CSSE). It contains country and state-wise daily new cases, recovered and death data of COVID-19.

%% file: figs2/wordcloud.tex
\begin{figure*}[h!]
    \centering
    \begin{subfigure}{0.48\textwidth}
        \centering
        \includegraphics[width =\textwidth]{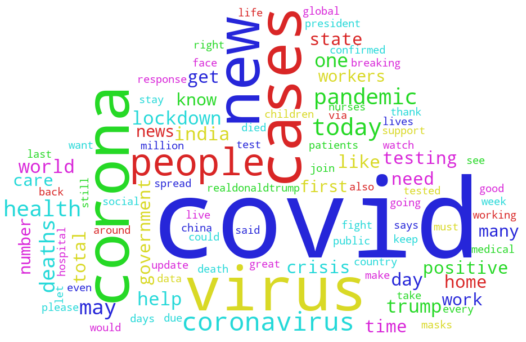}
        \caption{}
        \label{Full world}
    \end{subfigure}%
    ~ 
    \begin{subfigure}{0.48\textwidth}
        \centering
        \includegraphics[width =\textwidth]{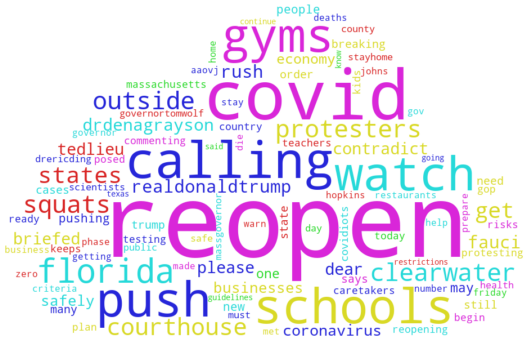}
        \caption{}
        \label{Only US-reopen}
    \end{subfigure}
    \caption{\textbf{Word Cloud Representations.} These are the visual representations of words within tweets whose importance is visualized by way of their size. (a) word cloud representation of the overall dataset, (b) word cloud representation of reopen related tweets within the United States.}
\end{figure*}

%% file: methodology.tex
\section{Methodology}
\label{sec:methodology}

In this section, we will describe the data preprocessing, and the experimental setup for sentiment and emotion analysis.

\subsection{Data Preprocessing}
\label{sec:methodology_preprocessing}
The data we used from the Twitter post between May 3-15, 2020 is noisy and unstructured text in nature. Therefore, we applied the following preprocessing steps to clean the raw tweet.
\smallskip
\begin{itemize} [nosep]
    \item URL and Email: Removing URLs and emails from tweets using  regular expression.
    \item Mention and Tagging: Removing mentions (@ symbol and word after @ symbol) and hashtags (only \# symbol) from tweets using regular expression.
    \item Noisy Words: Removing irrelevant words (i.e. RT, \&amp;) after manual inspection, and stripping non-ASCII characters from the tweet.
    \item Newline and Whitespace: Removing newlines and extra whitespaces from tweets.
    \item US-reopen: In section \ref{sec:result_most_freq} - \ref{sec:result_sentiment}, we first filtered tweets by locations (States of US) and then selected tweets having word `\textit{reopen}'.
    \item Stopwords and Gazetteers: In section \ref{sec:result_most_freq} for most frequent words, we removed words from tweets that are found in the Python NLTK \texttt{stopwords} and \texttt{gazetteers} corpus \footnote{\url{https://www.nltk.org/api/nltk.corpus.html}}. We also split the tweets into a list of words using the Python NLTK TweetTokenizer, and represent different forms of words to a single word using the Python NLTK WordNetLemmatizer ~\cite{NLTKToolKit}.
    \item Multi-line Tweet: In section \ref{sec:result_emotion} for emotion analysis, we converted the multi-line tweet into a single-line tweet in order to use the corresponding APIs.
\end{itemize}

\subsection{Sentiment Analysis}
We used the Python library \texttt{TextBlob} \cite{loria2018textblob} for finding the sentiments from tweets. In \texttt{TextBlob}, sentiments of tweets are analyzed in two perspectives: (1) Polarity and (2) Subjectivity. Based on tweets, this library returns the polarity score and subjectivity score. Polarity score is float value within the range [-1 to 1] where 0 indicates Neutral sentiments, the positive score represents Positive sentiments and the negative score represents Negative sentiments. Subjectivity score is also a float value within the range [0 to 1] where 0 indicates that it is a Fact and other values indicate the public opinions.

\subsection{Emotion Analysis}
\label{sec:methodology_emotion}
We used the \textit{Python} implementation of IBM Watson™ Tone Analyzer service \footnote{\url{https://cloud.ibm.com/apidocs/tone-analyzer?code=python}} to detect emotional tones in tweets. The \texttt{ToneAnalyses} API uses the general-purpose endpoint to analyze the emotional tones of a tweet and reports the following seven tones: analytical, tentative, confident, joy, fear, sadness, and anger. The emotion detection process is summarized as following steps: 
\smallskip
\begin{itemize} [nosep]
\item Create a Tone Analyzer service in IBM Cloud for a specific region and API plan.
\item Get the API key, the service endpoint URL, and the version of API.
\item Authenticate to the API using IBM Cloud Identity and Access Management (IAM).
\item Request up to $100$ sentences (no more than $128$ KB size in total) for the sentence-level tone analysis.
\item Parse the JSON responses that provide the results for each sentence of the tweet.
\end{itemize}
\smallskip
Here for our analysis, the \texttt{ToneAnalyses} requires a tweet as a single sentence for detecting sentence-level emotional tones.

%% file: results.tex
\section{Results and Analysis}
\label{sec:results}

In this paper, we seek to answer the following research questions.
\smallskip
\begin{itemize} [nosep]
    \item RQ1: What are the most frequent words?
    \item RQ2: What are the trends of sentiment?
    \item RQ3: What are the trends of emotion?
    \item RQ4: What are the effects of reopening?
\end{itemize}
\smallskip

\input{frequent}

\input{figs2/emotion}
\input{emotion}

\input{figs2/sentiment}

\input{sentiment}

\input{figs2/reopen}
\input{reopen}

%% file: frequent.tex
\subsection{RQ1: Most Frequent Words}
\label{sec:result_most_freq}
\input{wordcloud}

\input{ngrams}

%% file: wordcloud.tex
\subsubsection{Word Cloud Representation}

The tweets were organized into word clouds to analyze which words have been frequently used by Twitter users around the world.  As word cloud visualization consists of the size and visual emphasis of words being weighted by their frequency of occurrence in the textual corpus, we can get insights from the most frequent words occurring. From figure \ref{Full world}, we can see that along with the words `COVID', `corona', or `virus', words like `new', and `cases' got a large number of mentions. No wonder the rising number of new cases everyday around the world contribute to the good number of mentions of these words. Our main goal is to see the effect of reopening in our analysis that's why we also investigate the most frequent words used in reopen related tweets. Figure \ref{Only US-reopen} represents only the reopen related most frequent words within the United States. We can see that `businesses', `schools', `gyms', `economy', and `protesters' are the most frequently used words. Few tweets are `Protesting to reopen the economy', `reopen the economy against shelter in place which is a false choice', `As schools make plans to reopen after \#COVID19 shutdowns, how  should we proceed?', `Any rush to reopen without adequate testing and tracing will cause a resurgence.', `As we reopen shuttered offices and buildings, there is a danger of an outbreak of Legionnaire disease'. From these few tweets, we can see that few people are protesting to reopen, but most of the people are concerned about the reopening timing, aftermath of reopening, and thinking about the precaution measures needed after reopening.

%% file: ngrams.tex
\subsubsection{N-gram Representation}

We searched for the most frequent words that appeared in twitter posts of our collected data. For this section, we only considered actual tweets and ignored all kinds of re-tweets from our dataset. After applying the preprocessing steps mentioned in section \ref{sec:methodology_preprocessing}, we also removed hashtags and non-alphabetic characters from tweets. We split the rest of the tweet into chunks of $n$ consecutive words. Next, we merged the chunks of all tweets as a flat list, where $n \in [1,2,3]$. After that, we count the frequency of each unique chunk and sorted those by frequency. The top $15$ most frequent chunks, known as $n$-gram, are shown in Table ~\ref{table:top_ngram}.

\begin{table} [h!]
    \begin{center}
        \resizebox{\columnwidth}{!}{%
        \begin{tabular}{|c|c|c|c|c|c|c|}
            \hline
            \textbf{1-gram} & \textbf{2-gram} & \textbf{3-gram} \\ \hline 
            \hline
            
            state & testing site & drivethru testing site \\ \hline
            business & social distancing & moderate discussion business  \\ \hline
            plan  & white house & discussion business leader \\ \hline
            economy & state begin & keep calling transparency \\ \hline
            people & small business & calling transparency investigation \\ \hline
            today  & look like & transparency investigation fort \\ \hline
            back & public health & stock fluctuated economic \\ \hline
            case & stay home & fluctuated economic plan \\ \hline
            testing & back work & economic plan mixed \\ \hline
            country & wear mask & plan mixed earnings \\ \hline
            need & hair salon & mixed earnings company \\ \hline
            phase & next week & training class course \\ \hline
            county & task force & class course offer \\ \hline
            restaurant & business begin & course offer full \\ \hline
            week & contact tracing & offer full service \\ \hline
            
        \end{tabular}%
        }
        \caption{Top-$15$ N-grams (US-reopen).}
        \label{table:top_ngram}
    \end{center}
\end{table}

From Table ~\ref{table:top_ngram}, we can see that the `business', `economy', `back', and `need' clearly indicate the economic breakdown and necessity to overcome this situation through reopen. The `plan', `case', `testing', and `phase' surely emphasize the safety measures that need to be followed while reopening as a control measure. The `state' and `people' surely stands for how all states beginning to reopen in some way.

From 2-grams, the `social distancing', `stay home', `wear mask', and `contact tracing' imply users try to raise awareness through their posts. The `testing site', `public health', and `task force' imply the safety measures people and policymakers should abide by to curb the spread during the reopen situation. The `state begin',  `small business', `back work', and `business begin' indicate as people are going to resume their work after states are reopening. 

From 3-grams, the economic breakdown, stock-market fluctuation, business plan, and training course are the topic that draws our attention. This surely indicates there were numerous posts about how the loss due to this pandemic can be overcome through the strategy of the people who are in charge of the decisions.

%% file: figs2/emotion.tex
\begin{figure*} [h!]
\noindent \begin{minipage}{.5\textwidth}
\includegraphics[width=0.99\linewidth]{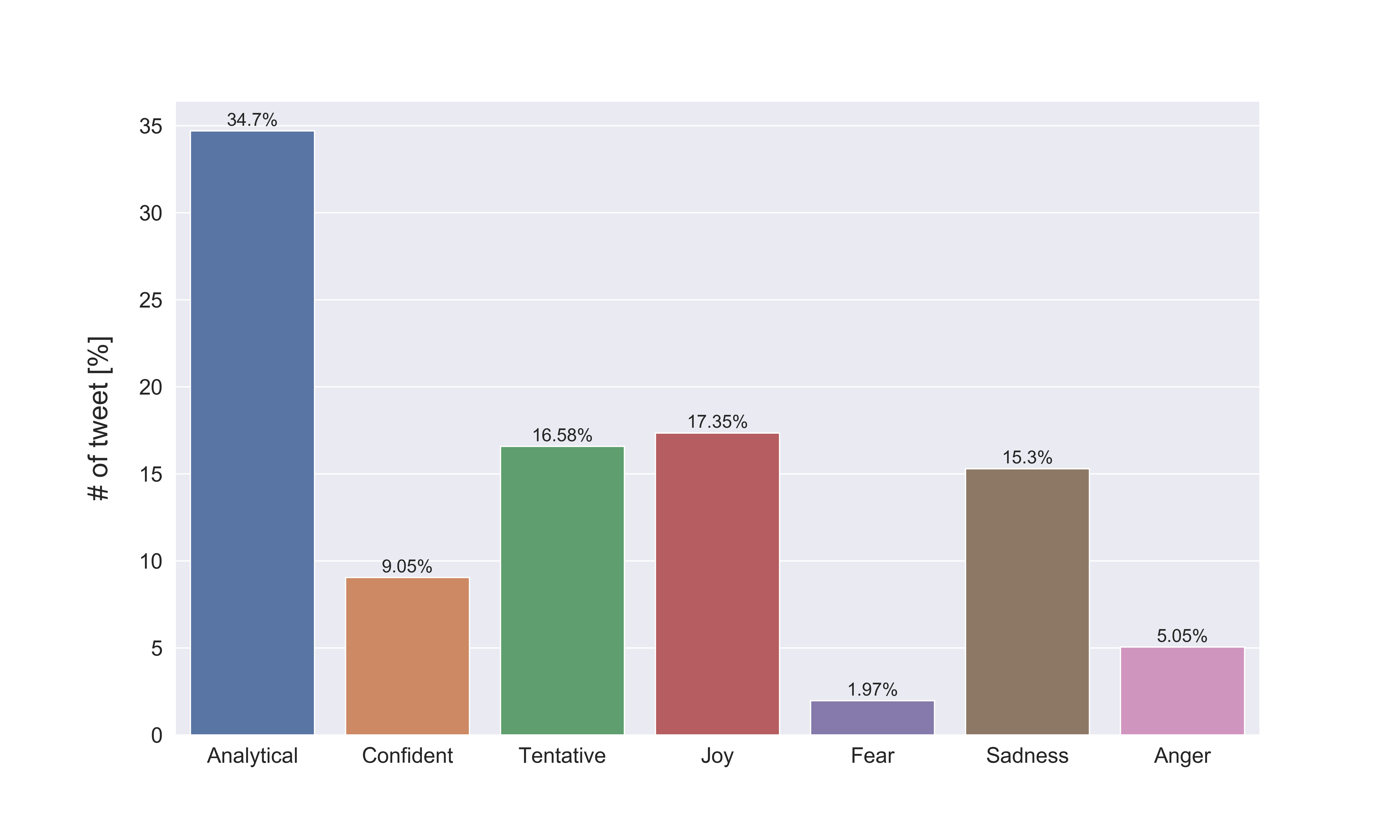}
\caption*{(a)}
\end{minipage}%
\begin{minipage}{.5\textwidth}
\includegraphics[width=0.99\linewidth]{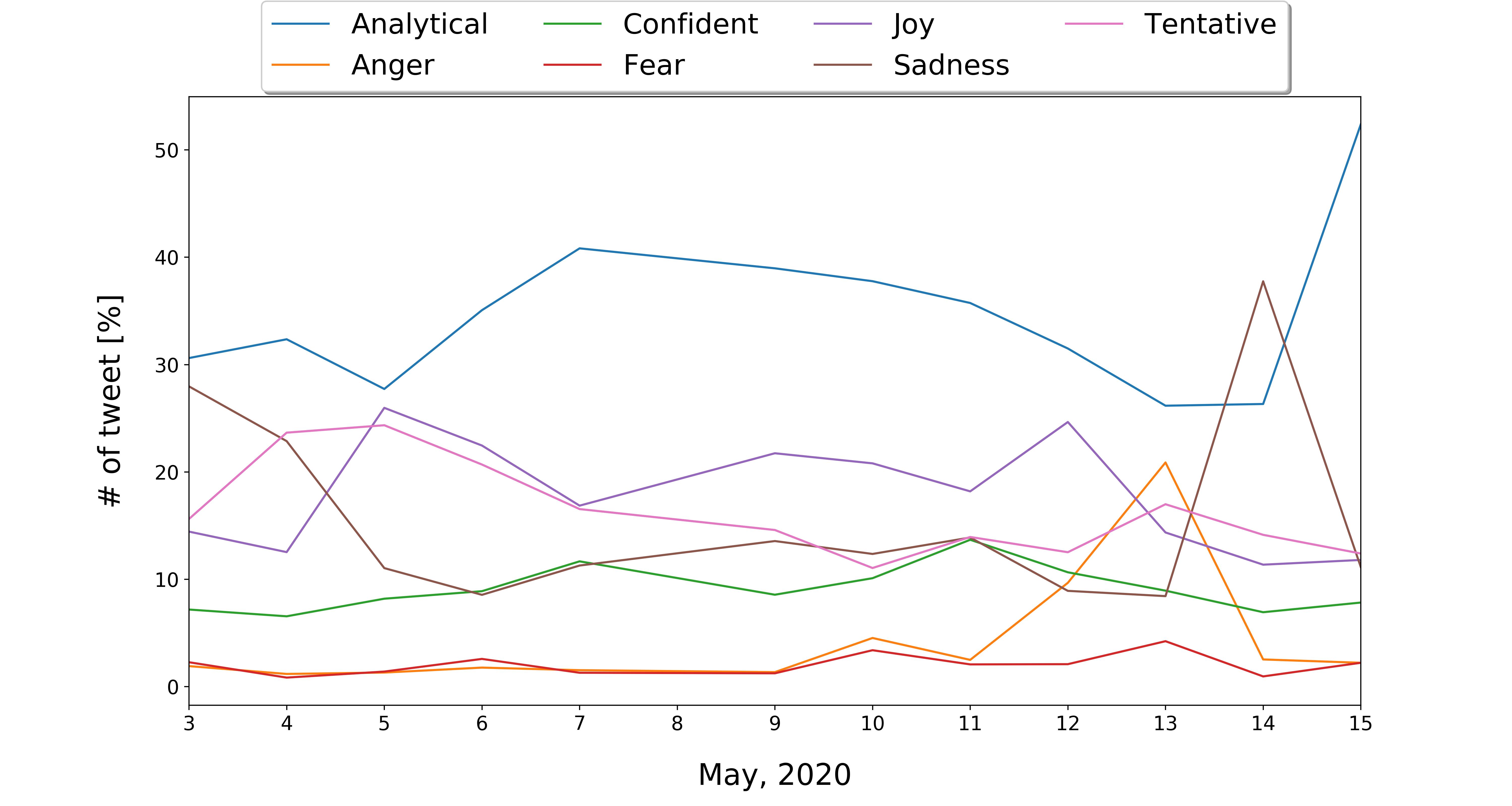}
\caption*{(b)}
\end{minipage}
\caption{\textbf{Emotional Distribution.} (a) tweet percentages of different kinds of emotional tones (b) daily tweet percentages of different kinds of emotional tones.}
\label{fig:emotion_perc_type}
\end{figure*}

%% file: emotion.tex
\subsection{RQ2: Emotion Analysis}
\label{sec:result_emotion}

From figure \ref{fig:emotion_perc_type}a, we can see that the highest percentage of emotional tone is `Analytical'(34.7\%). The second highest tone was `Joy'(17.35\%). The next few tones are `Tentative', `Sadness', and `Confident', respectively. `Anger' and `Fear' had the lowest percentage in our data collection.

The `Analytical' tone implies that people shared information and their constructive thoughts through their posts. `Joy' implies that people showed a positive attitude towards the situation. \cite{newazCovid} showed that towards the end of March the `Fear' emotion was the most dominant among all emotions and hit a peak. But after the announcement of the lockdown easement program people had less fear. 

From figure \ref{fig:emotion_perc_type}b, we can see that, during May 13-15, `Anger', `Sadness' and `Analytical' had high peaks. The rest emotional tones `Confident', `Joy', `Tentative', and `Fear' had a steady curve during the time period of our data collection.

%% file: figs2/sentiment.tex
\begin{figure*}[h!]
    \centering
    \begin{subfigure}{0.5\textwidth}
        \centering
        \includegraphics[width =\textwidth]{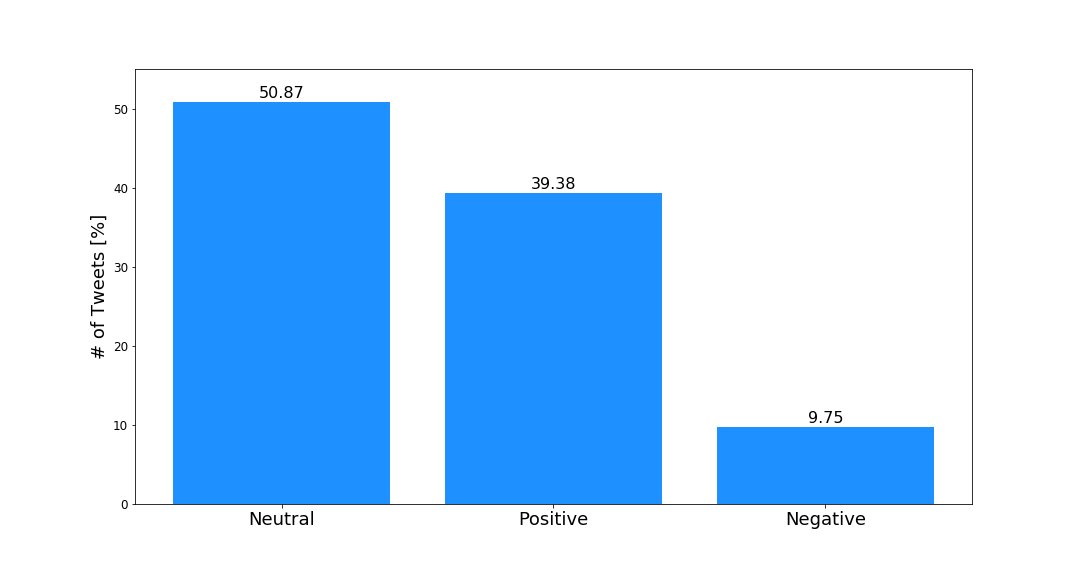}
        \caption{}
        \label{fig:positive_negative_barplot}
    \end{subfigure}%
    ~ 
    \begin{subfigure}{0.5\textwidth}
        \centering
        \includegraphics[width =\textwidth]{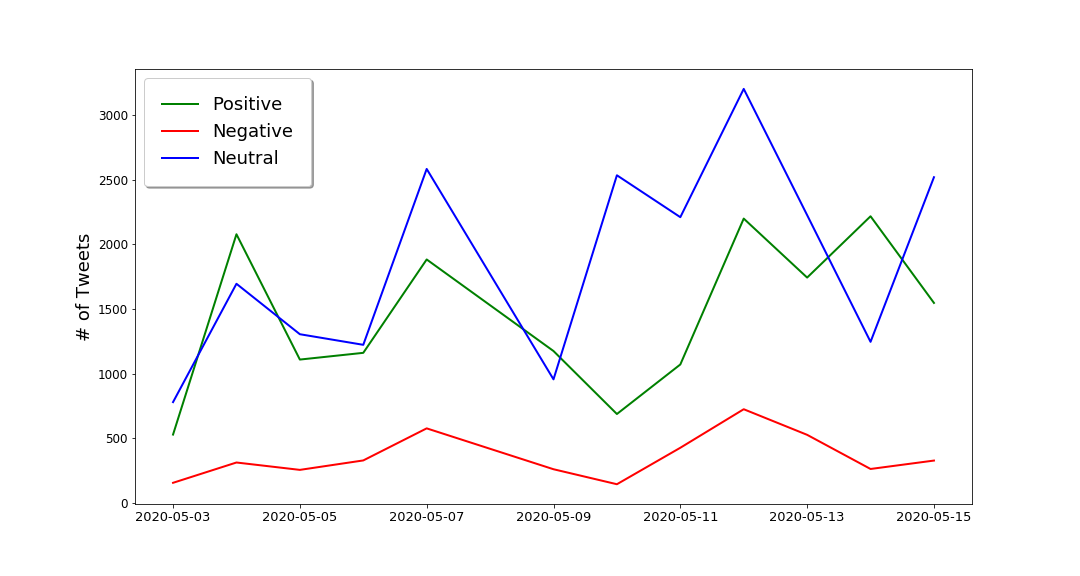}
        \caption{}
        \label{fig:positive_negative_lineplot}
    \end{subfigure}
    
    
    \begin{subfigure}{0.5\textwidth}
        \centering
        \includegraphics[width =\textwidth]{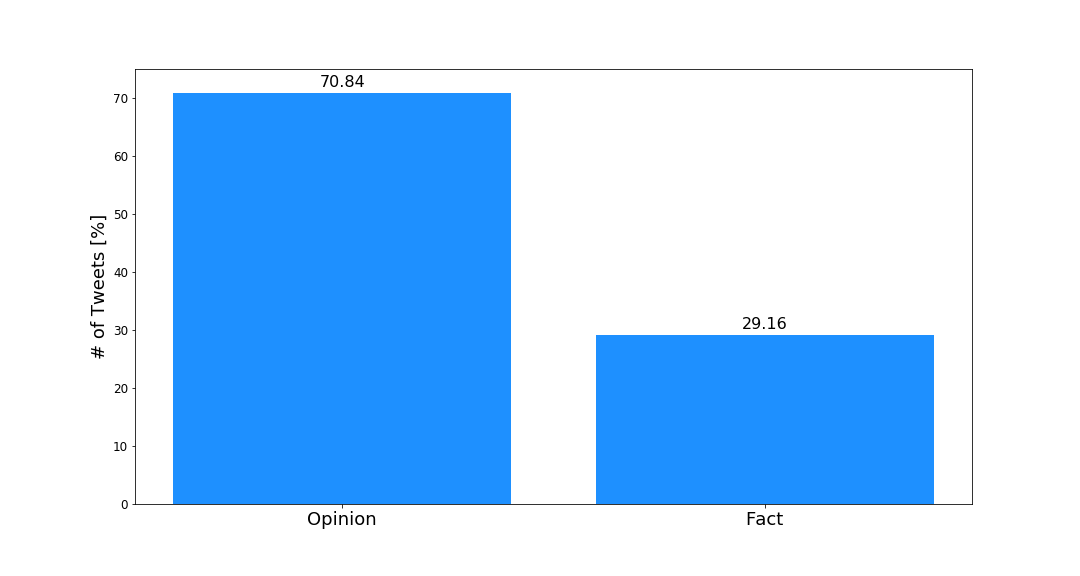}
        \caption{}
        \label{fig:subjectivity_barplot}
    \end{subfigure}%
    ~ 
    \begin{subfigure}{0.5\textwidth}
        \centering
        \includegraphics[width =\textwidth]{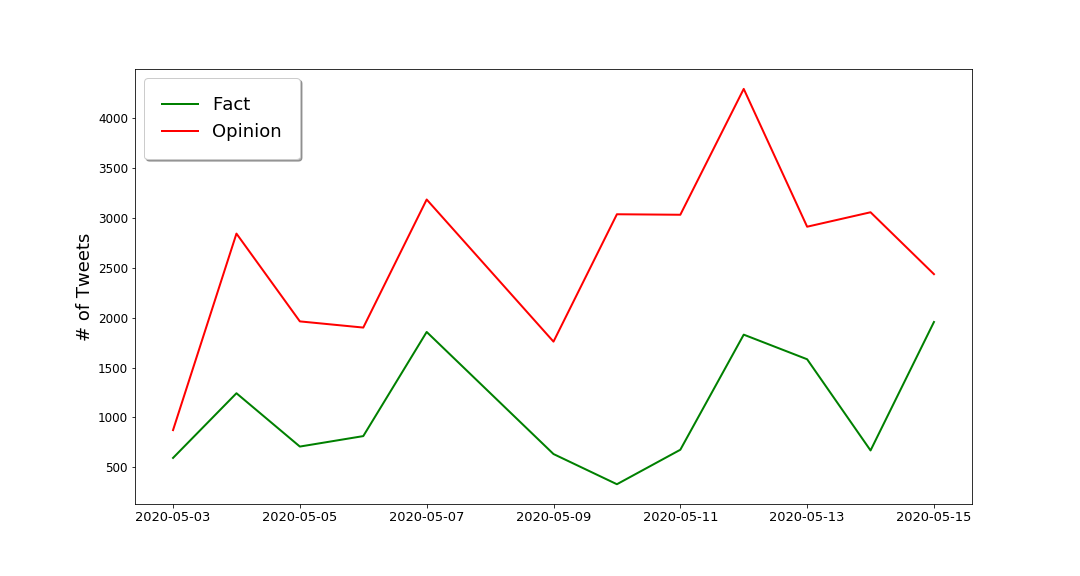}
        \caption{}
        \label{fig:subjectivity_lineplot}
    \end{subfigure}
    \caption{\textbf{Sentiment Distribution.} (a) tweet percentages of different kinds of sentiments (b) daily tweet counts of different kinds of sentiments (c) distribution of subjectivity types and (d) daily tweet counts of different kinds of subjectivity.}
    \label{fig:positive_negative_sentiment}
\end{figure*}

%% file: sentiment.tex
\subsection{RQ3: Sentiment Analysis}
\label{sec:result_sentiment}

Sentiment analysis identiﬁes words’ contextual polarity and subjectivity. We used Python library \texttt{TextBlob} \cite{loria2018textblob} for sentiment analysis. The details of \texttt{TextBlob} is available in section \ref{sec:methodology}. Fig. \ref{fig:positive_negative_sentiment} represents the details of twitter sentiments. Fig. \ref{fig:positive_negative_barplot} corresponds to the barplot of sentiment polarities of tweets and Fig. \ref{fig:positive_negative_lineplot} corresponds to the daily counts of polarities. It can be seen that the majority of tweets have a neutral sentiment (43.66\%) followed by positive (39.89\%) sentiments. Also, the daily trend of positive and negative tweets follows the same pattern. So, people are taking the reopen decision positively. Fig. \ref{fig:subjectivity_barplot} indicates that people are mostly posting their own opinion and Fig. \ref{fig:subjectivity_lineplot} shows an interesting point that day by day people are posting their own opinion rather than facts.

%% file: figs2/reopen.tex
\begin{figure*}[h!]
\includegraphics[width=0.7\textwidth]{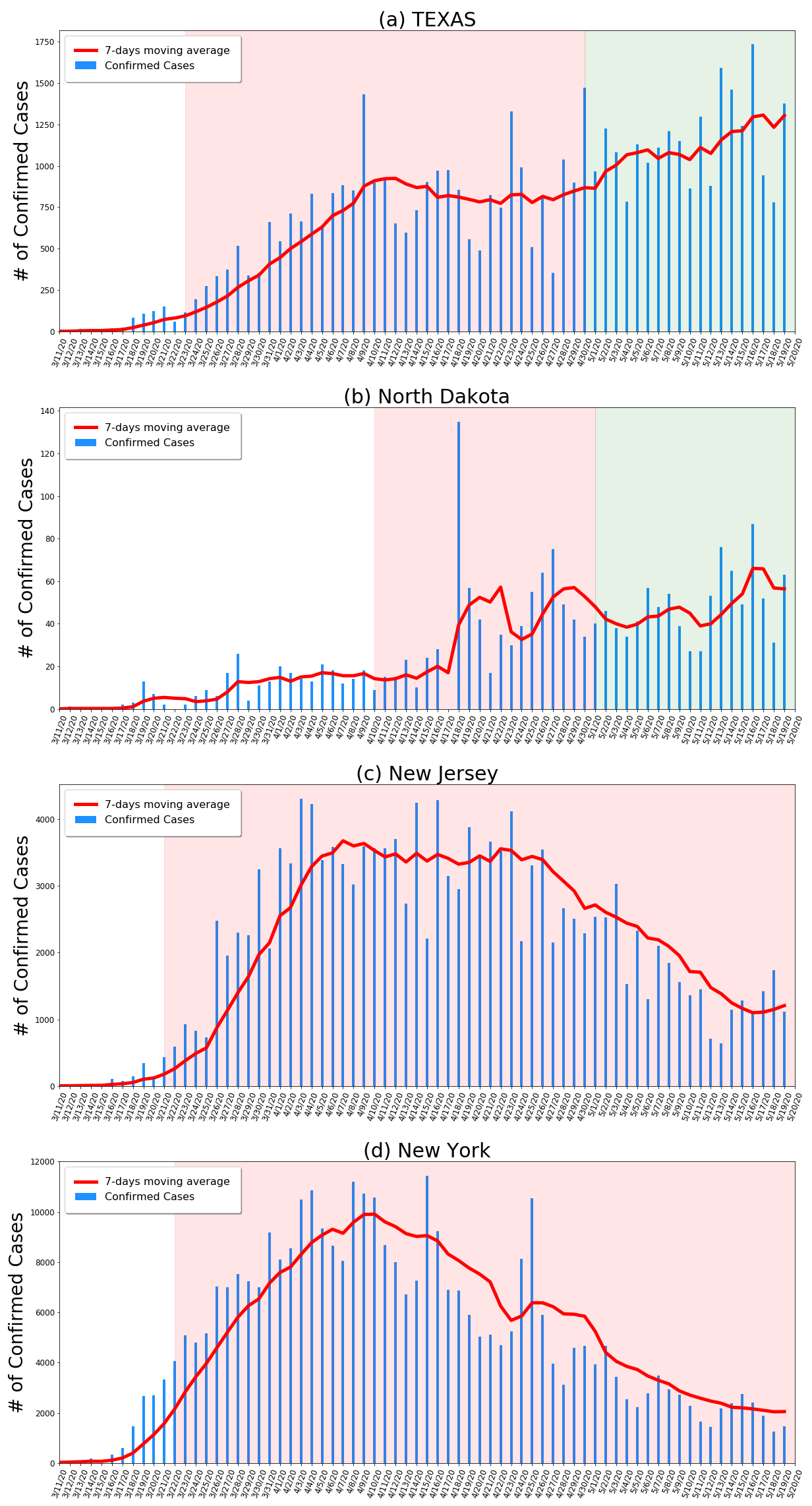}
\centering
\captionof{figure}{\textbf{Effect of reopen.} This figure represents the daily confirmed cases from 4 different states of the United States. We used two kinds of shaded regions to indicate the lockdown and reopen times: red shaded area represents the lockdown period, and the green shaded area represents the reopen period.}
\label{fig:reopen_effect}
\end{figure*}

%% file: reopen.tex
\subsection{RQ4: Effect of Reopen}
\label{sec:result_effect}

We saw that many peoples are protesting for reopening states and countries. We also saw it in Fig. \ref{Only US-reopen}. We investigated the reopening effect in the United States. Fig. \ref{fig:reopen_effect}(a) and (b), shows that after reopening the new COVID19 cases are increasing. On the other hand, Fig. \ref{fig:reopen_effect}(c) and (d), shows that lockdown helps to reduce the number of new COVID19 cases. Lockdown also helps to reduce the new cases in other countries \cite{lau2020positive}. It would be better if the lockdown continues for few more days or if the states or country decide to reopen, then they can take proper precautions before the reopen and people need to strictly maintain the instructions from health organizations.

%% file: future.tex
\section{Threats to Validity}
\label{sec:threats}
The followings are the main challenges and threats to the validity of our study.

\Part{Limited Data.}
We only conducted a shallow analysis of the twitter post from May 3,  2020 to May 15, 2020. Therefore, our results may not reflect the actual effect. We leave the large scale evaluation as future work.

\Part{Noisy Data.}
The dataset contains posts from social media (i.e. Twitter) that are noisy and unstructured.  Having mis-spelling and unknown words are also common in the post.  Additionally, the dataset contains other languages but written using the English alphabet that can negatively impact the analysis. There are many posts whose length is very small (i.e. around 3 words). Therefore, dataset cleaning is challenging and important as it has a direct impact on the analysis. 

\Part{Internal Validity}.
We used some open source toolchains in this paper and some issues may exist in those APIs. However, to reduce the issue in our implementation, other collaborators reviewed the code and manually inspected some results.

%% file: conclusion.tex
\section{Conclusion}
\label{sec:conclusion}
In this work, we try to get insights into public reaction as the reopen phase starts. There has been some analysis from social media data about how people are reacting in the lockdown time. We make an effort to understand whether there is a change in public sentiment from the lockdown phase to the reopen phase. We have made our analysis on Twitter data on reopening related issues during this COVID-19 outbreak. 
From our analysis, we have found that even though people are showing a positive attitude to reopen their areas to make the economy functional, they are also urging the authority in concern to make plans for avoiding the highly predictable second wave. From real data, we have seen that the new cases are increasing as the reopen phase starts. Given this situation, the coordination of responses from online and from the real world is sure to reveal promising results to address the current crisis. We want to put stress on the fact that along with traditional public health surveillance, infoveillance studies can play a vital role.

\smallskip
\Part{Source Code.} We have released the source code for public dissemination and can be found at \href{https://github.com/emtiaz-ahmed/COVID19-Twitter-Reopen}{https://github.com/emtiaz-ahmed/COVID19-Twitter-Reopen}.